\documentclass[aip,amsmath,amssymb,reprint]{revtex4-1}
\usepackage{graphicx}
\usepackage{dcolumn}
\usepackage{bm}
\usepackage[utf8]{inputenc}
\usepackage[T1]{fontenc}
\usepackage{mathptmx}

\begin{document}

\title{Terahertz charge and spin transport in metallic ferromagnets: the role of crystalline and magnetic order}
\author{Kumar Neeraj}
\affiliation{Department of Physics, Stockholm University, Stockholm, Sweden}
\author{Apoorva Sharma}
\affiliation{Institute of Physics, Chemnitz University of Technology, 09126 Chemnitz, Germany}
\author{Maria Almeida}
\affiliation{Center for Microtechnologies, Chemnitz University of Technology, 09126 Chemnitz, Germany}
\author{Patrick Matthes}
\affiliation{Fraunhofer Institute for Electronic Nanosystems, 09126 Chemnitz, Germany}
\author{Fabian Samad}
\affiliation{Institute of Ion Beam Physics and Materials Research, Helmholtz-Zentrum Dresden-Rossendorf, Dresden, Germany}
\author{Georgeta Salvan}
\affiliation{Institute of Physics, Chemnitz University of Technology, 09126 Chemnitz, Germany}
\author{Olav Hellwig}
\affiliation{Institute of Physics, Chemnitz University of Technology, 09126 Chemnitz, Germany}
\affiliation{Institute of Ion Beam Physics and Materials Research, Helmholtz-Zentrum Dresden-Rossendorf, Dresden, Germany}
\author{Stefano Bonetti}
\email[Corresponding author: ]{stefano.bonetti@fysik.su.se}
\affiliation{Department of Physics, Stockholm University, Stockholm, Sweden}
\affiliation{Department of Molecular Sciences and Nanosystems,
Ca' Foscari University of Venice, 30172 Venice, Italy}

\begin{abstract}
We study the charge and spin dependent scattering in a set of CoFeB thin films whose crystalline order is systematically enhanced and controlled by annealing at increasingly higher temperatures. Terahertz conductivity measurements reveal that charge transport closely follows the development of the crystalline phase, with increasing structural order leading to higher conductivity. The terahertz-induced ultrafast demagnetization, driven by spin-flip scattering mediated by the spin-orbit interaction, is measurable in the pristine amorphous sample and much reduced in the sample with highest crystalline order. Surprisingly, the largest demagnetization is observed at intermediate annealing temperatures, where the enhancement in spin-flip probability is not associated with an increased charge scattering. We are able to correlate the demagnetization amplitude with the magnitude of the in-plane magnetic anisotropy, which we characterize independently, suggesting a magnetoresistance-like description of the phenomenon.\end{abstract}

\maketitle

The pioneering work of Beaurepaire \textit{et.al}\cite{beaurepaire1996ultrafast}, where an ultrafast loss of magnetic order (i.e. demagnetization) induced by femtosecond near-infrared (NIR) laser pulses was observed, has paved the way for optical control and manipulation of spins at sub-picosecond time scales. Following this work, several other experiments confirmed the ability of photons to induce magnetization dynamics on such ultrafast time scales\cite{koopmans2000ultrafast, koopmans2005unifying, koopmans2010explaining,dalla2007influence}. This opened up the possibility to write magnetic information using light, with the prospect of a faster and more energy efficient storage technology\cite{stanciu2007all,khorsand2012role}. However, a full understanding of how ferromagnets dissipate their spin angular momentum at such short time scales is still lacking. One key aspect has been particularly complex to assess: the role of lattice and lattice dynamics. While some first studies addressing the coupling between phonon dynamics and ultrafast magnetism have appeared \cite{henighan2016generation,dornes2019ultrafast}, there are very few systematic studies on the role of the crystalline structure itself \cite{unikandanunni2021anisotropic}. A common intrinsic difficulty in these studies is the highly non-equilibrium state created by photons in the eV range, which often hides the dynamics occurring at the much lower energies proper of lattice excitations.

In recent years, however, the use of intense terahertz (THz) radiation to study ultrafast magnetization dynamics in various materials \cite{kampfrath2013resonant,shalaby2018coherent,shalaby2016low,bonetti2016thz,polley2018terahertz,polley2018thz,hafez2018extremely,hudl2019nonlinear,neeraj2021inertial,walowski2016perspective,seifert2018terahertz,vedmedenko20202020,kovalev2018selective} has gathered much attention, owing to the lower photon energy (in the meV range), much closer to the characteristic lattice energies. Indeed, early indications that the lattice structure plays a role in terahertz-driven demagnetization were already discussed in Ref. \cite{bonetti2016thz}. In that work, the authors measured two samples with completely different crystal structure, a crystalline iron and an amorphous CoFeB film. They reported that the presence of defect sites in the amorphous samples led to a larger demagnetization, due to enhanced spin-flip scattering at the defect sites which were in a much smaller number in the crystalline sample.

In this Letter, we design a study to investigate in greater detail how terahertz-driven demagnetization and crystalline structure are related. We choose a similar CoFeB amorphous system studied in Ref. \cite{bonetti2016thz}, but now we systematically change its lattice structure from amorphous to crystalline under controlled annealing, as shown in Ref. \cite{sharma2020crystallization}. This study shows that annealing increases the ordering of the lattice, monitored with X-ray diffraction, spectroscopic ellipsoemtry, and magneto-optical Kerr effect spectroscopy. In turn, the different lattice structure affects the optical and magneto-optical properties of the sample. Here, by comparing the ultrafast demagnetization driven by intense THz fields, and the THz electronic conductivity, we find indication of a more complex interplay between magnetic and structural order, where enhanced spin-dependent scattering is observed without an associated enhancement of the charge scattering at an intermediate state where the magnetic anisotropy is enhanced.

\begin{figure*}[ht!]
\includegraphics[width=1\textwidth]{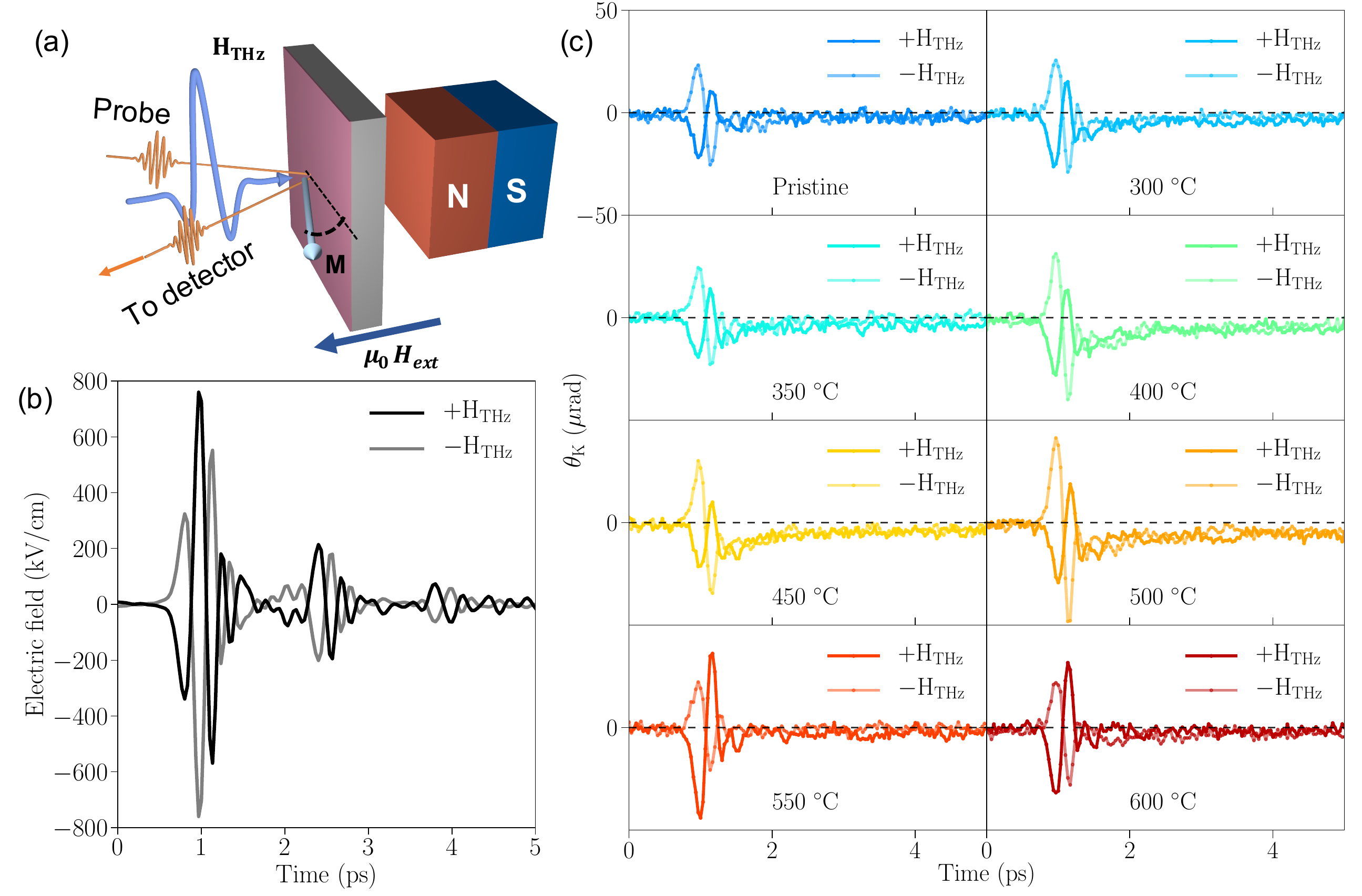}
\caption{(a) Schematic of the experimental geometry, with all relevant parameters discussed in the main text. (b) Measured terahertz electric field characterized via electro-optic sampling for the two opposite polarizations used in the experiments. (c) Measured polar MOKE for all samples with the corresponding annealing temperature indicated in the legend. All measurements were performed at room temperature, weeks after the anneling process was completed.}
\label{fig_1}
\end{figure*}

Pristine CoFeB films were deposited via magnetron sputtering \cite{sharma2020crystallization}. The stacking structure of our samples is Co$_{60}$Fe$_{20}$B$_{20}$ (100 nm)/Pt (5nm), grown on a 500 $\mu m$ thick silicon substrate with a native oxide layer on top. The samples were oven-annealed at different temperature ranging from 300 $^\circ$C to 600 $^\circ$C in steps of 50 $^\circ$C for 60 minutes. The samples were mounted in a THz-pump / NIR-probe setup, in the geometry shown in Fig. \ref{fig_1}(a), on top of a permanent magnet producing a magnetic field $\mu_0H\approx400$ mT. This magnetic field causes the easy plane magnetization vector \textbf{M} to tilt out-of plane, while preserving a sizable in-plane component. An intense THz magnetic field $\mathbf{H}_\mathrm{THz}$ with polarization normal to the direction of \textbf{M} excites the system which maximizes the $\mathbf{M}\times\mathbf{H}_\mathrm{THz}$ torque on the magnetization \cite{bonetti2016thz,hudl2019nonlinear,neeraj2021inertial}. The dynamical response of the magnetization is measured in the polar magneto-optical Kerr effect (MOKE) geometry using a nominally 40 fs near-infrared probe pulse and a balanced detection scheme \cite{hudl2019nonlinear}.

The intense THz fields are generated by optical rectification of a 1500 nm pulse (generated by optical parametric conversion from the 800 nm laser fundamental) in an organic DSTMS crystal \cite{puc2019dstms,jazbinsek2019organic}. By performing an electro-optic sampling of the THz pulse in a 50 $\mu$m thick GaP crystal \cite{hoffmann2011intense}, we reconstruct the temporal variation field of the single-cycle THz pulse shown in Fig. \ref{fig_1}(b). We estimate a free-space peak electric field of approximately 800 kV/cm, corresponding to a peak magnetic field of 270 mT, which lasts for less than a picosecond. In Fig. \ref{fig_1}(b) we also show that we can monitor and control the sign of the THz field (both electric and magnetic components) by rotating the THz-generating organic crystal by 180 degrees, producing the fields that we label $+H_{\rm THz}$ and $-H_{\rm THz}$ throughout this work. The signals detected at approximately 2.5 ps and 4 ps are the echos of the terahertz pulse traveling through the GaP crystal, reflected at each surface. They are shown for completeness, but they are not relevant for the actual experiment, since the much thicker silicon substrate separates these echos by more than 10 ps, beyond the measurement range of interest. For both the electro-optical sampling and the actual measurements, the time-delay between pump and probe beams is accurately controlled by a delay stage in one of the two beams' paths, given that both pulses originate from the same laser and are hence intrinsically synchronized.

Fig. \ref{fig_1}(c) shows the polar MOKE signal as a function of delay between pump and probe, mapping the magnetization dynamics initiated by terahertz pulses with opposite polarity. From these plots, we first notice the two distinct dynamics already observed in Refs. \cite{bonetti2016thz,hudl2019nonlinear}: a coherent response of the magnetization, which is associated with the torque exerted by  the magnetic component single-cycle THz field and which depends on its sign, and an incoherent response that is independent on the field polarity. The latter is understood in terms of an ultrafast demagnetization which is recovered within a few picoseconds. Furthermore, considering that all subplots were plotted on the same scale, and that the measurements were performed in the same geometry, we observe a variation of the maximum amplitude of the MOKE signal between samples that were annealed at different temperatures. In particular, the MOKE response has a relatively low peak-to-peak amplitude (50 $\mu$rad) in the pristine sample, which reaches a maximum of 100 $\mu$rad at an annealing temperature of 500 $^\circ$C, and then decreases at even larger annealing temperatures. In the Supplementary Material, we calculate this variation more precisely by taking the maximum difference, i.e. max$[\theta_K(+H_{\rm THz})-\theta_K(-H_{\rm THz})]$ between the MOKE responses to THz fields with opposite polarity, hence removing the one-sided offset caused by the demagnetization. With this analysis, we find that the amplitude variation of the coherent response is consistent with what was observed in the previous studies on similar samples \cite{hoffmann2019spectroscopic, sharma2020crystallization}. This observation is very important, because it provides us with a normalizing method within the data themselves, allowing us to separate magneto-optical effects from intrinsic magnetic ones, which are the ones of interest in our work.

\begin{figure}[t]
\includegraphics[width=0.48\textwidth]{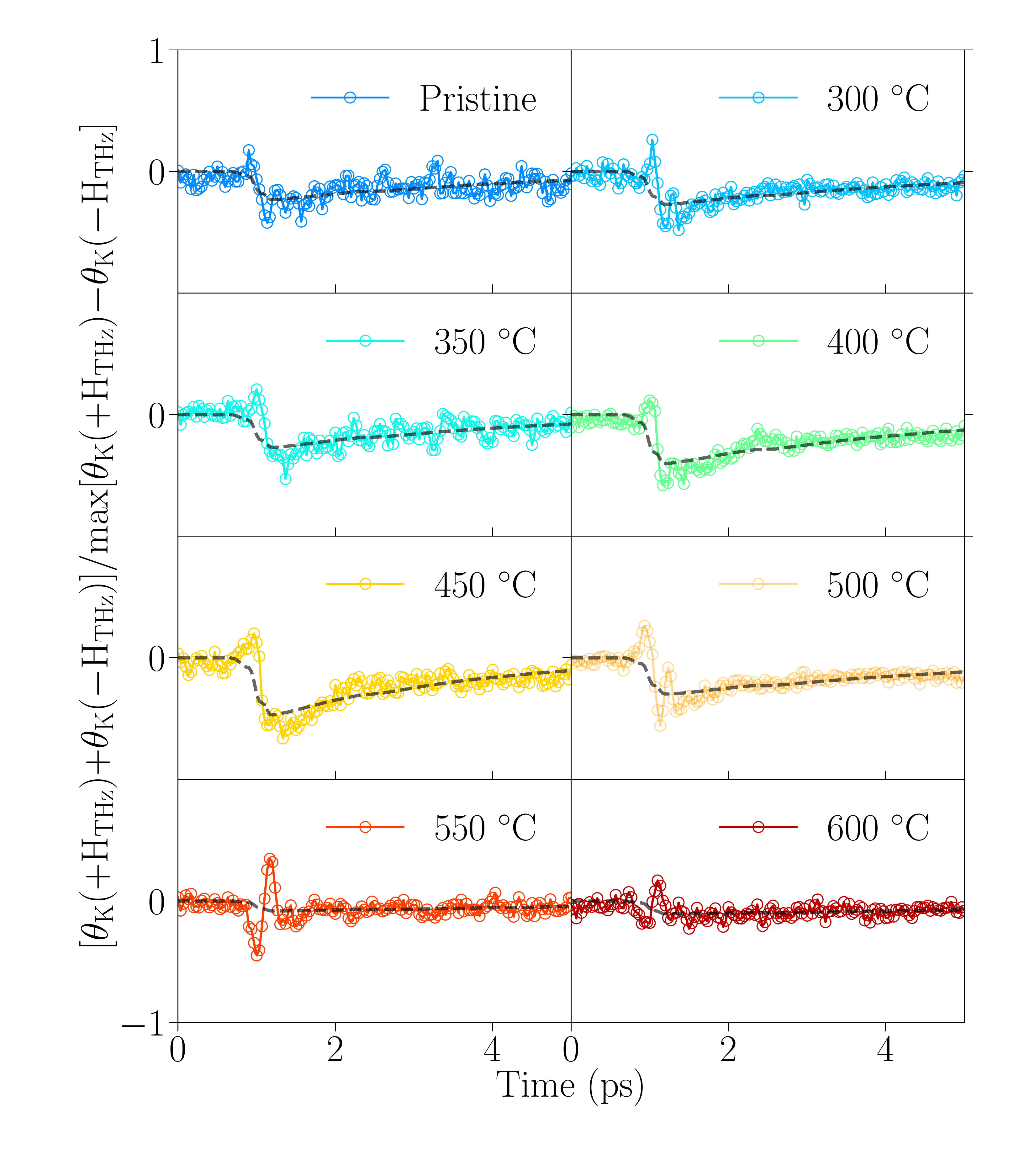}
\caption{Relative variation of the terahertz-induced demagnetization in samples with different annealing temperature, indicated in the respective legend. The axes are the same for all subplots. The dashed black gray line is a fitting to the data with the function given in Ref. \cite{bonetti2016thz}.}
\label{fig_2}
\end{figure}
We now turn to the incoherent response of the sample to the terahertz field, i.e. to the ultrafast demagnetization driven by the spin-polarized current flowing uniformly in the sample, which dissipates the energy deposited by the pump pulse \cite{bonetti2016thz}. In order to better separate the demagnetization from the coherent response, we follow the procedure explained in Ref. \cite{bonetti2016thz, hudl2019nonlinear}: in Fig.~\ref{fig_2}, we plot the sum of the MOKE signal originating from THz fields with opposite polarities, i.e. $\theta_K(+H_{\rm THz})+\theta_K(-H_{\rm THz})$. This procedure cancels out most of the coherent dynamics and allows us to perform a fit to the data using the approach of Ref. \cite{bonetti2016thz}. Since the MOKE signal varies with annealing temperature, we also normalize the data by the maximum of the coherent MOKE response, as discussed above. Looking at the data in Fig. \ref{fig_2}, we can evince that the qualitative shape of the demagnetization is the same for all samples. However, we see an evident trend in the demagnetization amplitude: the effect is relatively small for the pristine sample and for those annealed at 300~$^\circ$C and 350~$^\circ$C, it then starts to grow and reaches a maximum at 450 $^\circ$C, to almost disappear at the highest annealing temperatures. We analyze the results of the fitting procedure in detail later in the text.

\begin{figure}[t]
\includegraphics[width=0.45\textwidth]{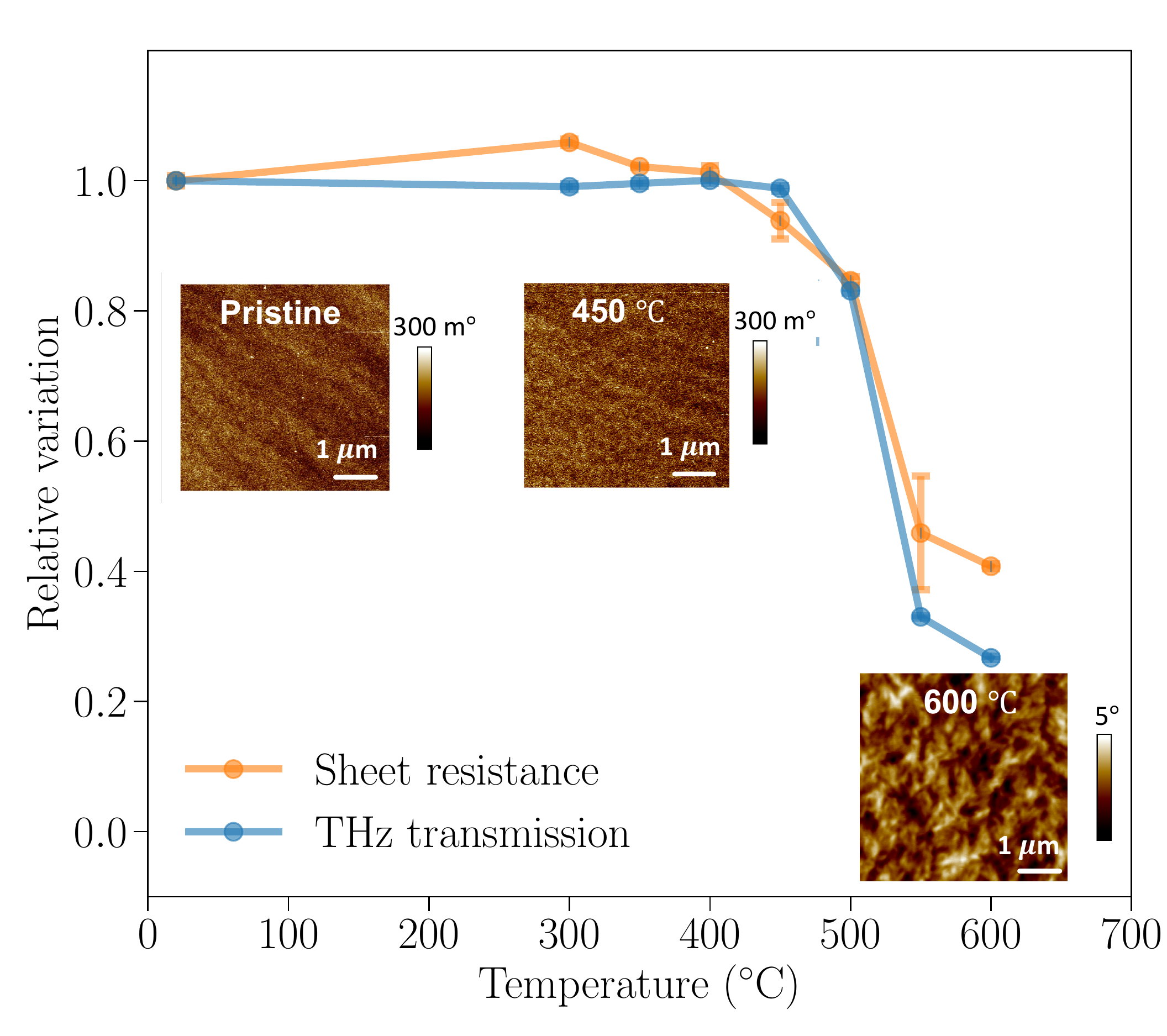}
\caption{Normalized sheet resistance and terahertz transmission as a function of annealing temperature for all films investigated in this work. The error bars in the THz transmission are calculated with the the transmission data available over a spectral range of approximately 2 THz, see Supplementary Material for details. Insets: selected MFM images for the pristine, 450 $^\circ$C annealed and 600 $^\circ$C annealed samples. The color bar close to each inset the MFM phase shift..}
\label{fig_3}
\end{figure}

Before moving to discuss the results so far obtained, we also estimate the resistivity variation between the different films using both conventional sheet resistance methods, as well as terahertz time-domain spectroscopy. The resistivity is a measure to quantify the \emph{charge} transport, with the terahertz data accessing the relevant demagnetization time scales. In metals the charge transport is strongly connected with the spin transport, and it is therefore an important parameter to estimate. We show in the Supplementary Material the details of the terahertz setup. In short, since the substrate is the same for all samples, the modulus of the relative THz transmission $|T|$ is proportional to the resistivity. In fact, from the Tinkham formula \cite{tu2003optical,laman2008terahertz} it follows that $\sigma=[(n_s+1)/(Z_0d)](1/|T|-1)\sim1/|T|$, where $n_s$ is the refractive index of the substrate, $Z_0=377$ $\Omega$ is the vacuum impedance, and $d$ is the film thickness. Hence, $\rho\sim|T|$. 
Fig. \ref{fig_3} shows both the sheet resistance and the THz transmission normalized to the value measured for the pristine samples. We notice that both measurements return values that remain more or less constant and similar to that of the pristine samples for all samples annealed up to 450 $^\circ$C, and then suddenly drop to less then half for the samples with the two highest annealing temperatures. In the insets, we plot the magnetic force microscopy (MFM) images for a few selected annealing temperatures, the full data is available in the Supplementary Material. They show that for an annealing temperature of up to 500 $^\circ$C, the magnetic contrast is rather smooth, with the tip phase shift being less than a degree. The MFM contrast changes abruptly for the 550 $^\circ$C (shown in the Supplementary Material) and 600 $^\circ$C annealed samples, with a maximum contrast more than one order of magnitude larger and with a clear evidence of a crystallization process which has taken place. This is consistent with the data from Ref. \cite{sharma2020crystallization}, where X-ray diffraction measurements on the same annealed samples as the ones used in this work showed a clear [110] CoFe peak rapidly forming as soon as the annealing temperature overcomes 500 $^\circ$C.


\begin{figure}[t!]
\includegraphics[width=0.5\textwidth]{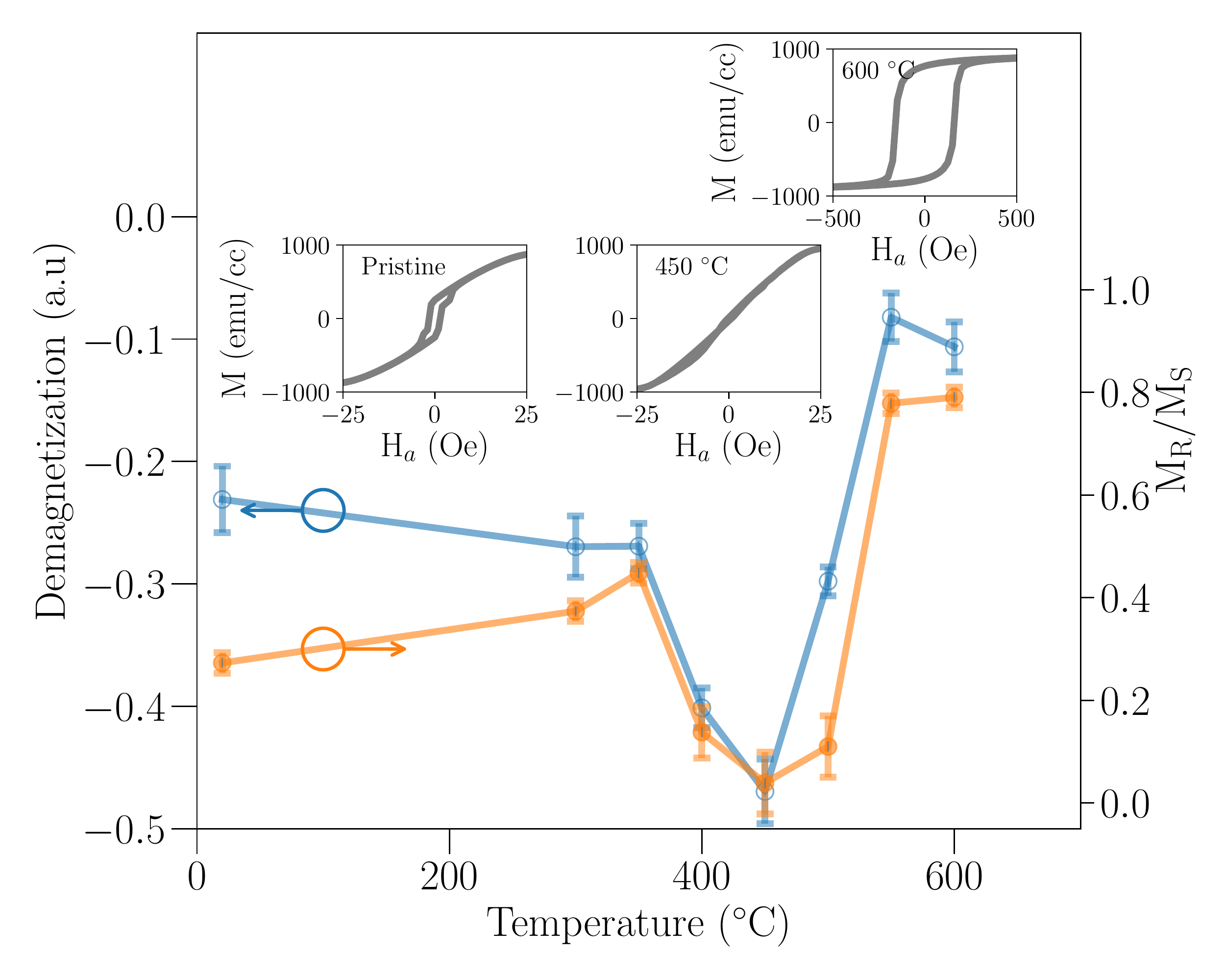}
\caption{Maximum demagnetization (blue) and relative remanence (orange) for samples annealed at different temperatures. The error bars in the demagnetization are calculated from the function used to fit the delay traces in Fig. \ref{fig_2}. Insets: hysteresis loops from the VSM measurements taken along the in-plane hard magnetization axis for the samples for a few selected annealing temperatures.}
\label{fig_4}
\end{figure}

In order to single out the change in magnetic properties from the crystalline ones, we measured the hysteresis loops for all samples by carefully rotating them in the sample plane in order to find the easy and hard magnetization axes. All hysteresis loops are reported in the Supplementary Material, and we plot a few characteristic ones along the in-plane hard magnetization axis in the insets of Fig.~\ref{fig_4}. For the pristine sample, there is a small remanence which is preserved for annealing temperatures up to 500 $^\circ$C, where the remanence and the coercivity go to zero. At higher annealing temperatures, the magnetic anisotropy is drastically modified: there are no clear in-plane hard and easy magnetization, with the two orthogonal directions having both a large coercivity and remanence. Plotting the normalized remanence for all samples, and overlaying it with the respective demagnetization amplitude, returns a remarkable observation: the demagnetization (i.e. the spin-flip probability of electrons accelerated by the terahertz field) correlates with the in-plane magnetic anisotropy of the sample. This evidence offers a novel insight into the ultrafast spin-flip process induced by terahertz radiation first reported in Ref. \cite{bonetti2016thz}. While it confirms the original observation that spin-flip scattering is enhanced in amorphous samples compared to crystalline ones, it also offers a deeper view of the mechanism. In particular, the new evidence of our work points to the fact that it is not only the larger number of defects (as in the case of an amorphous sample) that enhances the spin-flip probability, but also the strength of the spin-orbit coupling in the material. This confirms the hypothesis that the microscopic mechanism of terahertz-driven demagnetization is a spin-flip scattering of Eliot-Yafet type, proportional to the strength of the spin-orbit interaction. In fact, as shown in Fig. \ref{fig_3}, up to an annealing temperature of 450 $^\circ$C, the samples have approximately the same crystalline structure, which is not affecting the charge transport, resulting in a comparable resistance change. However, when the magnetic anisotropy changes in the film plane, the spin scattering apparently decouples from the charge one, and an enhanced demagnetization is observed. The entire scattering process can be seen in analogy with the well-known anisotropic magnetoresistance effect, where the relative change in resistance is of the order of 0.1\%, below the detection sensitivity of our resistivity measurements. However, the demagnetization experiments and the magneto-optical detection are only sensitive to the spin-dependent part of the scattering, and are able to clearly single it out.

In conclusion, we performed a systematic study on a series of CoFeB thin film samples which were grown under the same conditions to achieve an amorphous state, and which were then annealed at different temperaturesto induce an increasing degree of crystallization. We performed terahertz-induced ultrafast demagnetization and terahertz conductivity experiments to quantify the effect of annealing on both charge and spin scattering. We observed a decrease of the terahertz resistivity as soon as the annealing temperature induced a structural phase transition. The demagnetization experiments revealed a novel enhanced spin-flip process in an intermediate magnetic state characterized by the same electrical conductivity as the pristine samples, but with a much larger in-plane anisotropy. Such anisotropy must be due to an enhanced spin-orbit coupling in the sample plane caused by the annealing process, and is the microscopic mechanism that increases the spin-flip probability at crystalline defects. Our observations, supported by independent experimental data, further deepen our understanding of the fundamentals of spin and charge transport at ultrafast time scales, and their role in ultrafast magnetism of metals.

See the Supplementary Material for details on the time-resolved magneto-optical Kerr effect pump-probe setup, the analysis of the coherent MOKE response, the fitting model for the demagnetization data, the details on the terahertz time-domain spectroscopy setup and analysis, the vibrating sample magnetometer characterization of all samples, and for the full set of magnetic force microscopy images.

K.N. and S.B. acknowledge support from the European Research Council, Starting Grant 715452 MAGNETIC-SPEED-LIMIT. A.S., M.A. and G.S. aknowledge support from the Deutsche Forschungsgemeinschaft through the project 282193534 (Mechanisms of crystallization of CoFeB-based TMR stacks under laser annealing). P.M. acknowledges financial support through the European Regional Development Fund (EFRE) and the Free State of Saxony/ Germany (VP 3675, HZwo:FRAME VP2.2).

\section*{Data availability}
The data that support the findings of this study are available from the corresponding author upon reasonable request.

\end{document}